%% file: main.tex
\documentclass[12pt]{article}
\usepackage{amssymb}
\usepackage{amsxtra}
\usepackage{amsmath}
\usepackage{amstext}
\usepackage{amsthm}
\usepackage{amsbsy}
\usepackage{latexsym}
\usepackage{graphicx}
\usepackage{float}
\usepackage{xcolor}
\usepackage[font=small,labelfont=bf]{caption}

\usepackage{authblk}

\usepackage[utf8]{inputenc}
\usepackage{amsfonts}
\usepackage{hyperref}
\usepackage{lineno}
\usepackage[hscale=0.75,vscale=0.8]{geometry}
\usepackage{longtable}

\usepackage[compress]{natbib}

\begin{document}

\title{Pair density Shannon information measures and their $N$-dependent behaviour in diatomic molecular systems}

\author{Sa\'ul J. C. Salazar$^{1}$\thanks{{\tt col539936@colaborador.buap.mx}}, J. Antonio Z\'arate$^{2}$, J. M. Solano-Altamirano$^{1}$ \thanks{{\tt jmanuel.solano@correo.buap.mx}}, Humberto G. Laguna$^{2}$, Julio M. Hernández-Pérez$^{1}$ and Robin P. Sagar$^{2}$}
\affil{$^{1}$Facultad de Ciencias Qu\'imicas, Benem\'erita Universidad Aut\'onoma de Puebla, 14 Sur y Av. San Claudio, C.P. 72570 Puebla, Pue., M\'exico}
\affil{$^{2}$Departamento de Qu\'imica, Universidad Aut\'onoma Metropolitana Iztapalapa, Av. Ferrocarril San Rafael Atlixco 186, Col. Leyes de Reforma 1a Secci\'on, Iztapalapa, 09340 Ciudad de M\'exico}

\date{\today}

\maketitle

\begin{abstract}
Pair density Shannon entropies and mutual information in position and in momentum space are calculated for two series of homonuclear and heteronuclear diatomic molecules using Hartree-Fock wave functions. The pair density entropy sum increases with the quality of the basis set. Mutual information, a measure of statistical correlation, is seen to be smaller in momentum space as compared to position space. The interpretation is that the momentum pair density  is closer than the position pair density to a Hartree-like reference. The $N$-dependent behaviours of the entropy and mutual information sums are examined by fitting the data to three different model behaviours. Results show that these models are capable of representing the molecular data in differing degrees. The parameters obtained from a particular model are relatively constant across different chemical series. This adds evidence to the argument of a universal $N$-dependent behaviour of the entropy sums. 
\end{abstract}

\section{Introduction}

 Recent decades have witnessed an explosion of interest in the application of information theoretical ideas to examine the nature and behaviour of quantum systems. This has led to the establishment of quantum information theory as the framework for such studies. For the most part, attention has focused on discrete systems (e.g. spin) and the role that they could play in the development and construction of quantum technologies (e.g. quantum computers). This does not imply  that there is no interest in continuous variable systems. The majority of the interest here to date lies in the quantum optics community and indeed there are proposals for photonic quantum computers based on continuous variable systems. The focus here is examination and control of the behaviour in both the position and momentum representations, or indeed a combined phase-space one. A rough analogy between the two points of view is contrasting the information in a digital signal (discrete) with an analog (continuous) one. At present, there are also proposals for hybrid (discrete-continuous) quantum computing. Moreover, there is also keen interest from the confined quantum systems community in the application of information theory to understand the behaviour of systems such as quantum dots \cite{sen,nascimento,muker-roy1,sun,jiao,adan2019,nascimento2021,majumdar,mondal2022,kumar,estanon2023,chakladar}. Information theoretical measures have also been used in the study of Bose-Einstein condensates \cite{barnali1,barnali2,zhao}.

Interest in information theoretical measures has also been generated in the quantum chemistry community. There is current interest in the relation between quantum information theory and quantum chemistry \cite{boguslawski2013,aliverti2024,zhao2025}. Emphasis has been placed on the Shannon information entropy of the reduced one-electron density since this is a functional of the electron density, with its obvious connection and interest for the density functional theory communities. However there has also been interest in the Shannon entropy of the momentum space density. Efforts have been devoted towards the development of a momentum space quantum chemistry  \cite{thakkaracp}.

A principal reason for the interest in the Shannon entropies of both representations is that there is an entropic formulation of the uncertainty principle \cite{beckner1975,bbm1975,hertz2019}  which for atomic and molecular systems is 
\begin{equation}
S_t=S_{\rho}+S_{\pi} \ge 3(1+\ln \pi). 
\end{equation}
Multiplication by three in the above is due to the dimensions of the underlying densities. The Shannon entropies are
\begin{equation}\label{eq:shanonentropdef}
S_{\rho}=-\int \rho ({\bf r}) \ln \rho ({\bf r}) d{\bf r},  \qquad S_{\pi}=-\int \pi ({\bf p}) \ln \pi ({\bf p}) d{\bf p},
\end{equation}
where $\rho ({\bf r})$ and $\pi ({\bf p})$ are the respective one-electron reduced densities in position and in momentum spaces that are normalized to unity. 
The conventional manner to address the issue of the dimensional units of $\rho ({\bf r})$ and $\pi ({\bf p})$ inside the logarithm \cite{leipnik} is to divide the density inside the logarithm by a unity-valued reference density with the same dimensional units.

Entropic uncertainty relations are superior to variance based measures, especially when applied to systems with non-Gaussian behaviours such as atoms or molecules. They are of interest and used in fields such as quantum thermodynamics and continuous variable quantum information. The entropy
sum thus provides a basis independent measure of uncertainty.

There is also a link between the entropy sum and phase-space formulations. This lies in the consideration of this sum as an entropy of a separable phase-space distribution, 
\begin{equation}
S_t=  -\int \rho ({\bf r})\pi ({\bf p}) \ln \left[\rho ({\bf r})\pi ({\bf p})\right] d{\bf r} d{\bf p}.
\label{phase-space}
\end{equation}
This quantity is known as the Leipnik entropy \cite{leipnik}.

$S_{\rho}$, $S_{\pi}$ and $S_t$ have been examined in atomic and molecular systems \cite{gadre,hocpl,chatzi,hazra}. $S_t$ has been proposed and studied as a measure of wave function quality and seen to increase in value with better basis sets. It has also been examined in relation to electron correlation effects \cite{guevara2003pra}. Indeed, the position space Shannon entropy has been linked to the correlation energy in the electron gas \cite{grassi}.
This entropy has also been shown to be related to the logarithmic mean excitation energy, a quantity accessible from stopping power experiments \cite{ho-pra-1998}. The impulse behind these works is the quest for an understanding of what the entropies of electron densities represent, and the nature of how they quantify the information present in a chemical system. If the entropy sum is used, then one can argue in terms of a more complete idea of quantum information in the system.

On the other hand, there have been a few works on information measures for the pair densities of atomic systems \cite{sagar2011ijqc,lrosaijqc,torresijqc} and even fewer on molecular systems \cite{ayers2025}. There is current interest in the information theoretic approach in density functional theory \cite{liu1,liu2,liu3,liu4}. The motivation for such endeavours lies in the belief that the pair density contains more information that the single particle one. For example, information about chemical bonding and its different environments. This is partly responsible for the motivation behind a pair density functional theory. 

There is also an uncertainty entropic relation for the pair densities \cite{bbm1975,yanez,guevara2003jcp}. It is 
\begin{equation}
S_T=S_{\Gamma}+S_{\Pi} \ge 6(1 + \ln \pi),
\end{equation}
where the pair Shannon entropies are 
\begin{equation}\label{eq:pairentropiesdef}
S_{\Gamma}=-\int \Gamma ({\bf r_1,r_2}) \ln \Gamma ({\bf r_1,r_2}) d{\bf r_1}d{\bf r_2},  \qquad S_{\Pi}=-\int \Pi ({\bf p_1,p_2}) \ln \Pi ({\bf p_1,p_2}) d{\bf p_1}d{\bf p_2},
\end{equation}
with $\Gamma ({\bf r_1,r_2})$ and $\Pi ({\bf p_1,p_2})$ the respective pair densities normalized to unity. In the same way as the reduced density, the pair density within the logarithm is divided by a reference density with a unit value and the same dimensional units.
The interpretation of these entropies is that they are global measures of the delocalization (localization) in the underlying pair densities. Larger values correspond to a more delocalized distribution. One of the goals of this work will be to examine the behaviour of $S_T$ to see if there is indeed an increase with a better quality basis set. 

Another attractive feature of calculating pair Shannon entropies is that they offer a path to the calculation of mutual information. This quantity is a measure of the statistical correlation between variables and is defined in each representation as 
\begin{gather}
I_r=\int \Gamma ({\bf r_1,r_2}) \ln \left[\frac{\Gamma ({\bf r_1,r_2})}{\rho ({\bf r_1})\rho ({\bf r_2})}\right] d{\bf r_1}d{\bf r_2}=2S_{\rho}-S_{\Gamma} \ge 0, \\
 I_p=\int \Pi ({\bf p_1,p_2}) \ln \left[\frac{\Pi ({\bf p_1,p_2})}{\pi ({\bf p_1})\pi ({\bf p_2})}\right] d{\bf p_1}d{\bf p_2}=2S_{\pi}-S_{\Pi} \ge 0.
\end{gather}
It can be considered as the relative entropy between non-separable (numerator of logarithmic argument) and separable (denominator) pair densities in each space. The densities in the denominator of the logarithmic argument 
are reduced ones or marginals of the pair density.
That is, they are obtained by integration of the parent pair density over one of the variables, 
\begin{equation}\label{eq:marginals}
\rho({\bf r})=\int \Gamma ({\bf r_1,r_2}) d{\bf r_{2}},  \qquad \pi({\bf p})=\int \Pi ({\bf p_1,p_2}) d{\bf p_{2}}.
\end{equation}
The reduced densities here, obtained by integration over particle one or two, are the same since the particles are indistinguishable. This is emphasized by dropping the subscript on the first variable.
Mutual information is zero (uncorrelated) when the density in the numerator of the logarithmic argument is equal to the density in the denominator. The mutual information sum is defined as $I_r+I_p$ and will be the focus in this study.

The work will focus on calculations obtained from the Hartree-Fock (HF) scheme. The HF pair density is 
\begin{gather}
\Gamma_{HF}({\bf r_1,r_2})=\frac{1}{N-1}\left[N\rho ({\bf r_1})\rho ({\bf r_2}) -\Gamma _x ({\bf r_1, r_2}) \right], \\
\Pi_{HF}({\bf p_1,p_2})=\frac{1}{N-1}\left[N\pi ({\bf p_1})\pi ({\bf p_2}) -\Pi _x ({\bf p_1, p_2}) \right],
\end{gather}
where the $\Gamma _x ({\bf r_1, r_2})$ and $\Pi _x ({\bf p_1, p_2})$ second terms are the Hartree-Fock exchange densities in position and momentum space. These exchange densities are normalized to unity. For a pair density consisting of only the (separable) Hartree first term, $S_{\Gamma}$ and $S_{\Pi}$ are exactly twice the corresponding one-electron reduced Shannon entropies, $S_{\rho}$ and $S_{\pi}$. It is the exchange density in the HF scheme which is responsible for a non-zero mutual information. In this respect, mutual information can be thought of as representing the distance from a separable Hartree-like pair density. This informational distance is due to the effects of the exchange density. Smaller values are thus related with more Hartree-like behaviour. One can also examine the Shannon entropy values to gauge the extent of localization/delocalization presented by the Fermi hole which will also be a theme of this work. 

One can see from the relations that mutual information cannot be negative-valued because the influence of the exchange density and the Fermi hole is to localize the distribution as compared to the Hartree-like reference. Thus the localization features of $S_{\Gamma}$ and $S_{\Pi}$ offer insights into the influence of the exchange density and Fermi hole in a particular system.

The HF wave function is not a separable one since it comes from a single determinant. Correlation here is a result of particle indistinguishability. There is an active discussion at present about the role of indistinguishability and its relation to quantum entanglement. Here, many emphasize the need to consider multi determinant wave functions in discussions of entanglement.

$I_r$ and $I_p$ have been studied in chemical systems \cite{sagar2005jcp1,sagar2006jcp2,angulo2022,schurger2023,schurger2024}. 
A further goal will be a comparison of the magnitudes of these two measures in molecular systems. It has been reported that $I_r > I_p$ in atomic systems \cite{sagar2011ijqc,angulo2022}. 
The interpretation of this is that the pair density in momentum space is closer to the Hartree-like reference as compared to the pair density in position space and its reference. This suggests a kind of condensation in momentum space for these electronic (fermionic) systems. We will examine this relation here in molecular systems. 

There is another aspect of the information measures which we wish to discuss here. 
Movement away from the uncertainty bound is related with a loss of information as the entropy sum increases.
Understanding the $N$-dependent behaviour of the information measures and movement away from the bound, would help to establish the physical characteristics of information in chemical species. The $N$-dependencies of the Shannon entropy and mutual information sums are thus important as they provide information on how uncertainties and correlations behave with system size. 

It has been proposed \cite{gadre} that $S_t$ behaves as $a+b\ln N$ in neutral atoms. This behaviour was  noted in the Thomas-Fermi model \cite{gadrepra}. Such  behaviour was examined from calculations at the Hartree-Fock level with Slater type basis sets. This was further corroborated for other fermionic and bosonic type systems \cite{massenpla96,massenpla02,panos}. Other models for the $N$-dependent behaviour of entropy sums in different quantum systems have also been proposed \cite{guevara2003pra,salazar-physscr}. Another goal of this work is to examine these behaviours in molecular diatomic systems which to our knowledge have not been explored. These works have for the most part examined the behaviour of $S_t$, the one-electron reduced Shannon entropy sum. We will also examine the adherence of the pair entropy and mutual information sums to such behaviours in both neutral atoms and diatomic molecules.

\color{black}

\section{Results and Discussion} 

All informational entropies were computed with a developer version of DensToolKit \cite{solano2024} (see also Secs.~\ref{sec:compdutet} and \ref{sec:numintschs} for further details), using wavefunctions calculated with Gaussian 09 package \cite{frisch2009}. 
For diatomic molecules, we used the ground state configurations and bond distances provided by Huber \textit{et al.} \cite{huber1979} and Hostutler \textit{et al.} \cite{Hostutler}.

Tables 1 and 2 contain the values of the Shannon pair entropies and mutual information calculated from different basis sets at the Hartree-Fock level. The respective sums are also included. There are some general trends which can be observed from the data.
$S_T$ increases on going from the $STO-3G$ basis set to the others with one exception. 
This result is consistent with that of $S_t$ where it was proposed that $S_t$ is a measure of wave function quality \cite{gadre,hocpl}. Here, $S_{\Gamma}$ increases (delocalization) while $S_{\Pi}$ decreases (localization).


\input{tab-hxseries-hf-diffbasis}
\input{tab-xxneut-hf-diffbasis}

Note that  $I_r > I_p$ for all the diatomic systems. The interpretation here is that the exchange component of the pair density is relatively larger in position space as compared to momentum space. That is, in momentum space the pair density is closer to the Hartree-like reference. This result is in agreement with the results for atoms \cite{sagar2011ijqc,angulo2022}. 

 In the $X-X$ series, one can see $S_T$ increases with $N$. This is so because for larger $N$ more reductions have to be performed in order to reduce to the pair density. These integrations yield increasing uncertainty which translates into larger entropic values. One can see this from the $H_2$ results. Of all systems, it has the smallest $S_T$ values and lies closest to the bound since no reductions are performed to obtain the pair density of this two-electron system. It does not lie on the bound since it is not a Gaussian-type state (it is an exponential-type expanded into gaussian-type functions). Its mutual information in both representations is zero-valued because its pair density does not have an exchange component since the spins of the two electrons are different and the exchange density vanishes upon integration over the spins. 

On the other hand, $I_r$ and $I_p$ decrease with $N$ in the $X-X$ series. There is also a general trend in $H-X$ series. Thus, $I_t$ decreases with $N$ in both series.

\subsection{$N$-dependent functional relations of entropy sums} 

The $N$-dependent functional behaviours of entropies and entropy sums have been of interest in past years. 
Here, we will focus on the entropy sums due to their appearance in the entropic uncertainty relations.
In particular, Gadre obtained the relation 
\begin{equation}
S_t=a +b\ln N, 
\label{gadre-st}
\end{equation}
from consideration of the Thomas-Fermi model \cite{gadrepra}. This relation was later tested and found to hold for the electrons in neutral atoms \cite{gadre}. Furthermore, this behaviour has also been observed in other fermionic and bosonic systems, which has led to the supposition that it might be a universal behaviour obeyed by all particles \cite{massenpla96,massenpla02}. Indeed, this relation was used and argued \cite{panos} to examine
 the theory of gravity as an emergent phenomena \cite{verlinde}. 

Other models of entropy sum behaviour have also been presented. One such relation involves the dependence on the first ionization potential \cite{guevara2003pra} 
\begin{equation}
S_t=a\ln I_1 +b\ln N +c,
\label{ent-n-ion}
\end{equation}
and was obtained from modelling the asymptotic behaviour in the position space density and the cusp condition in momentum space \cite{guevara2003pra}. This hydrogenic model has been tested in the neutral atoms, where it was shown that the dependence on $I_1$ allows it to correctly reproduce the jumps  observed in the entropy sum with the appearance of a new shell. The Thomas-Fermi model is unable to reproduce these shell effects. 

A third model comes from consideration of the symmetric ground state of $N$-interacting one-dimensional harmonic oscillators \cite{salazar-physscr}. The entropy sum is given as
\begin{equation}
		S^{k}_{T}=k(1+\ln \pi) + \frac{1}{2}\ln \bigg[\frac{[\omega(N-k)+k\Lambda_{N}][\omega k+(N-k)\Lambda_{N}]}{N^{2}\omega \Lambda_{N}} \bigg], 
\label{stoscil}
	\end{equation} 
where $k$ is the reduction order. $k=1$ for $S_t$ while $k=2$ for $S_T$. $\omega$ is the intensity of the one-body harmonic potential while $\Lambda_{N}=\sqrt{\omega^{2}\pm N\lambda^{2}}$ contains $\lambda$, the intensity of the two-body harmonic potential. The $\pm$ in the square root defines the presence of an attractive ($+$) or repulsive ($-$) interaction potential. Note that the second term vanishes in a non-interacting system when $\lambda=0$. Thus, interactions are necessary in order to provoke dependencies from the second term. Eq. (\ref{stoscil}) also illustrates how the reduced entropy sums depend on the potentials in these harmonic oscillator systems.

This expression serves as a basis for understanding the origin of the parameters in the $a+b\ln N$ relation. One can see that the first term in Eq. (\ref{stoscil}) is the uncertainty bound and can be associated with $a$. Note that this term has to be multiplied by a factor of three to account for the dimensionality in the atomic and molecular problem here. 

The argument of the logarithm in the second term of Eq. (\ref{stoscil}) can be rearranged in terms of $N$ as
\begin{multline}
    1-\frac{2k}{N}+\frac{2k^{2}}{N^{2}}+\frac{k(N-k)}{N^{2}} \left ( \frac{\omega}{\Lambda_{N}}+\frac{\Lambda_{N}}{\omega} \right) = \\
    1 - \frac{k}{N} \Bigg[ 2- \bigg( \frac{\omega}{\Lambda_{N}} + \frac{\Lambda_{N}}{\omega} \Bigg) \Bigg] + \frac{k^{2}}{N^{2}}\Bigg[ 2 - \Bigg( \frac{\omega}{\Lambda_{N}} + \frac{\Lambda_{N}}{\omega} \Bigg) \Bigg].
\end{multline}
Thus, one can appreciate the logarithmic dependence on $N$ from the argument of the second term. However, the $N$-dependency is more complex than in Eq. (\ref{gadre-st}) since there are contributions from the magnitudes of both the one and two-body potentials in the argument. In this light, the $b$ parameter in Eq. (\ref{gadre-st}) models these complex dependencies. 

Comparisons can also be made to the second model in Eq. (\ref{ent-n-ion}). The $c$ parameter can be equated with the uncertainty bound while the first two terms can be combined into $\ln [I_1^{a}N^b]$. In this way, the logarithmic argument would have dependencies on both $N$ and the first ionization potential. However, the difference with Eq. (\ref{stoscil}) is that the potentials present are the one and two-body potentials and not the first ionization one. Relations between the mutual information sum and a logarithmic interaction energy in oscillators have been explored \cite{salazar-pre}.  


We will now inquire if Eq. (\ref{stoscil}) provides a useful representation for the information measure sums. To model the chemical behaviour present in atoms and molecules, we substitute $\omega$ by $N^{\alpha}$ to take into account the effects of the nuclear charge ($Z=N$) in an atomic system. $\alpha$ is introduced into the expression for flexibility to take into account the successive reductions needed from the $N$-particle density. We found that using a positive sign in the square root expression for $\Lambda _N$ provided a better representation. A factor of three was also included in the first term or uncertainty bound to take into account the dimensionality of the problem. The value of $k$ was set at unity for the $S_t$ fits and at two for $S_T$. 

Our goal here is an examination of the three models that will be fitted to the data of actual atomic and molecular systems. While the first two relations have been tested in neutral atoms, the last one has not. Moreover, neither of the three have been applied to diatomic molecular systems.  This is a major aim. Lastly, the first two measures have been proposed for $S_t$. That is, the entropy sum from the reduced one-electron densities. We will also probe their applicability (with the third relation) to $S_T$, the entropy sum for the pair densities. 

\begin{figure}[H]
		\includegraphics[width=0.97\linewidth]
		{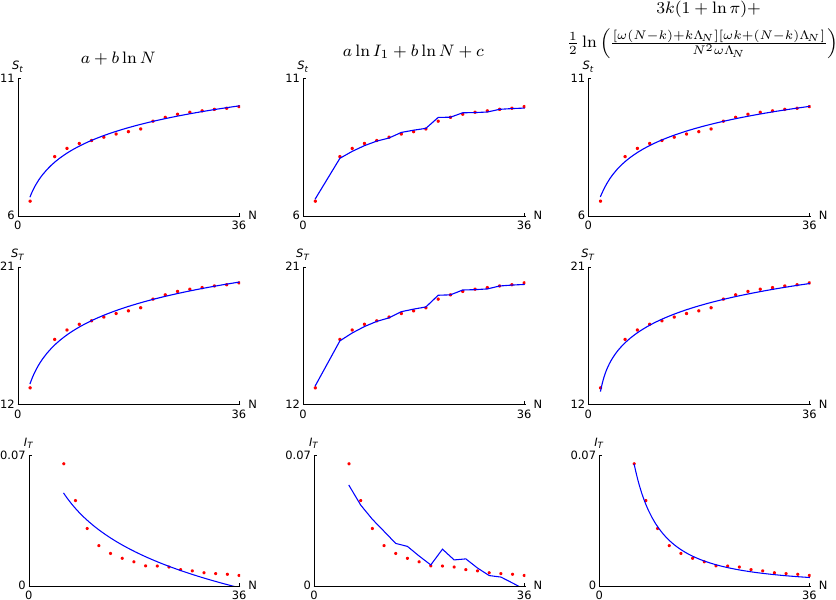}
		\caption{Shannon entropy sums $S_t$ (first row), $S_T$ (second row) and $I_T$ (third row) as a function of $N$ for the homonuclear diatomic molecules. Columns correspond to the first [Eq. (\ref{gadre-st})], second [Eq. (\ref{ent-n-ion})] and third [Eq. (\ref{stoscil})] models respectively. The red dots are the actual data points obtained with the 6-31 G basis set while the blue continuous curves are the respective fits. }
		\label{fig1}
	\end{figure}

Figure \ref{fig1} presents the data points along with the obtained fitted curves for the homonuclear diatomic series.  The information is arranged so that the columns contain the results for the three different models while the rows contain the different information measures. It is striking that all three models are able to represent the data points in this diatomic series. The plots in the middle column have more structure as compared to those in the other two columns. This is due to the dependence on the first ionization potential in this particular model. The other two models do not possess explicit dependencies on the first ionization potential. The values of $S_{\rho}$ and $S_{\pi}$ in the three series are not reported here. These are available upon request from the authors.

On the other hand, one can see that the mutual information sum is accurately represented by only the third model. This can be expected since the first two models do not explicitly contain information about pair densities while the third model does so. In  fact, it is also noteworthy that expressions for the first two models represent $S_T$ since these models were not derived with explicit information about pair densities. 

The corresponding Figures \ref{fig4} and \ref{fig5} related to the behaviour of the neutral atoms and heteronuclear diatomic molecules are presented in the Sec. \ref{sec:appendixc}. Both figures display the same tendencies that were discussed for the homonuclear diatomics. 

A more detailed comparison of the models can be performed by examining the actual parameters of the fits that were performed.
We report the values of adjusted coefficient of determination ($R^{2}_{adj}$) because it penalizes the inclusion of additional parameters, and allows a fairer comparison of models with different numbers of parameters than simply using $R^{2}$. Furthermore, to compare across models (linear models with different numbers of parameters and the nonlinear model), we used the corrected Akaike Information Criterion ($AICc$ \cite{akaike}), which is preferable when the sample size $n$ is small. A lower value indicates a better model (a balance between goodness of fit and complexity). Therefore, the model with the lowest $AICc$ value is preferred. That is, the model with a more negative or less positive value. This indicates a very good fit of the model to the data.
These metrics are presented in the following three tables where the information for all three chemical series  are presented for a particular model. This facilitates the analysis among series. 

\begin{table}[H]
	\begin{center}
		\begin{tabular}{c c c c c} 
			\hline
			$S_{t}$ & $a$ & $b$ & $R^{2}_{adj}$ & $AICc$ \\ [0.5ex] 
			\hline
			$\mbox {Neutral atoms (He-Ar)}$ & 6.2193 & 1.0775 & 0.9817 & -50.5908 \\ 
			$\mbox {Homonuclear molecules (X-X)}$ & 5.9105 & 1.1431 & 0.9839 & -21.0444 \\
			$\mbox {Heteronuclear molecules (H-X)}$ & 5.8795 & 1.2046 & 0.9748 & -8.2701 \\
			\hline 
			$S_{T}$ & $a$ & $b$ & $R^{2}_{adj}$ & $AICc$ \\ [0.5ex] 
			\hline
			$\mbox {Neutral atoms (He-Ar)}$ & 12.2589 & 2.1974 & 0.9828 & -2.9821 \\ 
			$\mbox {Homonuclear molecules (X-X)}$ & 11.7736 & 2.2967 & 0.9858 & 0.5285 \\
			$\mbox {Heteronuclear molecules (H-X)}$ & 11.6993 & 2.4212 & 0.9795 & 4.7525 \\
			\hline
             $I_{T}$ & $a$ & $b$ & $R^{2}_{adj}$ & $AICc$ \\ [0.5ex] 
			\hline
			$\mbox {Neutral atoms (He-Ar)}$ & 0.1962 & -0.0592 & 0.5581 & -62.6091 \\ 
			$\mbox {Homonuclear molecules (X-X)}$ & 0.1001 & -0.0281 & 0.8322 & -108.3501 \\
			$\mbox {Heteronuclear molecules (H-X)}$ & 0.1482 & -0.0511 & 0.8408 & -55.4847 \\
			\hline
		\end{tabular}
\caption{Values of $a$ and $b$ parameters obtained from fitting entropy and mutual information sum data to the $a+ b\ln N$ functional form. Fits to $S_t$, $S_T$ and $I_T$ for the neutral atoms, homonuclear diatomic molecules and heteronuclear diatomic molecules are presented.}
	\end{center}
\end{table}
Table 3 presents the parameter values for the neutral atoms and diatomic molecules. These values  obtained for $S_t$ of the neutral atoms using the 6-31G basis set are very similar to those reported in \cite{gadre}, taking into account the different basis sets that are used ($N$-normalized entropies instead of unity normalized ones were used in that work). All fits are relatively good with $R^{2}_{adj} =0.98$. It is striking that the $S_t$ parameter values are very similar, taking into account that the three series represent different chemical environments. The $a$ values are around the uncertainty bound value of $3(1+\ln \pi) \approx 6.43$. The $b$ values are around unity for the neutral atoms but seem to increase when chemistry is introduced with the molecules.  

These same trends are also observed for the $S_T$ parameters. That is, $a$ is close in value to the respective uncertainty bound
while $b$ increases with chemistry. While the $a + b\ln N$ provides a close representation of the $N$-dependent behaviour of the entropy sums, it is an empirical expression with questions about the nature of the $a$ and $b$ parameters. We have seen how $a$ can be approximately equated to the values of the uncertainty bounds. However, questions remain about the nature of $b$. 

The $a+b\ln N$ form can also be used to fit the mutual information sum values. Note the negative value of $b$ in the three sequences which captures the decreasing tendency of $I_T$ with $N$. Furthermore the values of $a$ are very small here as one would expect from the previous table and the definition of $I_T$ ($2S_t-S_T$). The fits here are not as good as the entropy sum ones as seen from the $R^{2}_{adj}$ values. This quantifies the visual evidence seen in Figure \ref{fig1} and the Sec. \ref{sec:appendixc}.  

The fitting parameters from the second model in Eq. (\ref{ent-n-ion}) are presented in Table 4. First, one can appreciate that the fits of the entropy sums to this model yield $R^{2}_{adj}$ values that are very close to unity. Indeed, the values for both diatomic series are similar or slightly better than the ones for the neutral atoms. Moreover, all values are closer to unity as compared to those from the first model in Table 3. Care should be exercised in this interpretation since this model consists of three linear parameters while the first one only has two parameters.  The $R^2_{adj}$ values suggest that this second model, with the inclusion of experimental data, provides a closer representation of the $N$-dependence of the entropy sums in the three series. The values of $c$ for $S_t$ and $S_T$ are close to the respective bounds. 

On the other hand, the fits to the $I_T$ values are not as good as the ones from the entropy sum. This mimics the behaviour seen in the first model. The $b$ parameter is also negative-valued to account for the decaying behaviour of $I_T$. $c$ is also small and approaches zero.

\begin{table}[H]
\begin{center}
		\begin{tabular}{c c c c c c} 
			\hline
			$S_{t}$ & $a$ & $b$ & $c$ & $R^{2}_{adj}$ & $AICc$ \\ [0.5ex] 
			\hline
			$\mbox{Neutral atoms (He-Ar)}$ & -0.2051 & 1.0443 & 6.0861 & 0.9912 & -75.2179 \\
			$\mbox {Homonuclear molecules (X-X)}$ & -0.2021 & 1.1373 & 5.7241 & 0.9939 & -36.6539 \\
			$\mbox {Heteronuclear molecules (H-X)}$ & -0.2891 & 1.1825 & 5.6491 & 0.9889 & -16.3286 \\
			\hline 
			$S_{T}$ & $a$ & $b$ & c & $R^{2}_{adj}$ & $AICc$ \\ [0.5ex] 
			\hline
			$\mbox{Neutral atoms (He-Ar)}$ & -0.4267 & 2.1281 & 11.9814 & 0.9928 & -32.3851 \\
			$\mbox {Homonuclear molecules (X-X)}$ & -0.3859 & 2.2855 & 11.4176 & 0.9949 & -15.8625 \\
			$\mbox {Heteronuclear molecules (H-X)}$ & -0.5305 & 2.3805 & 11.2765 & 0.9914 & -3.8447 \\
             \hline
             $I_{T}$ & $a$ & $b$ & $c$ & $R^{2}_{adj}$ & $AICc$ \\ [0.5ex]
             \hline
             $\mbox{Neutral atoms (He-Ar)}$ & -0.0691 & -0.0621 & 0.1252 & 0.6729 & -74.9302 \\
             $\mbox {Homonuclear molecules (X-X)}$ & -0.0076 & -0.0269 & 0.0885 &  0.8617 & -110.4444 \\
             $\mbox {Heteronuclear molecules (H-X)}$ & -0.0158 & -0.0489 & 0.1281 & 0.8597 & -55.7507 \\
             \hline
		\end{tabular}
\caption{Values of $a$, $b$ and $c$ parameters obtained from fitting entropy and mutual information sum data to the $a \ln[I_{1}] + b \ln[N ] + c$ form. Fits to $S_t$, $S_T$ and $I_T$ for the neutral atoms, homonuclear diatomic molecules and heteronuclear diatomic molecules are presented. $I_1$ data correspond to NIST \cite{nist}.}
	\end{center}
    \end{table}

Details of the fits with values of the  $\alpha$ and $\lambda$ parameters from the third model are presented in Table 5. First, one can appreciate from the $R^{2}_{adj}$ values that Eq. (\ref{stoscil}) provides a very good representation of the data points across all chemical species. Second, there is a marked improvement of these values (closer to unity) as compared to the ones in Table 3 for the $a+b\ln N$ fits. This contrasts with the interpretation from the $AICc$ metric where the larger negative values in Table 4 imply that the second model offers a better fit of the data.

The values of $\alpha$ are similar across chemical species and present an inverse behaviour with $N$. This inverse dependence shows how reductions from the $N$-particle level influence the behaviour of the sums. The $\lambda$ values are positive valued and are largest for the neutral atoms. This parameter models the interaction among the particles. It is noteworthy that it is the parameter with the most variation across the chemical species. In this light, it captures the varying chemical environments across the different series.

A further point deserves discussion. Eq. (\ref{stoscil}) is an expression that pertains to interacting oscillator systems. Table 5 also shows that the fits provide a good description to chemical systems. However, the behaviour here for these systems must include a more complex dependence in the logarithmic argument than that provided by the oscillators in Eq. (\ref{stoscil}). Evidence for this point can be taken from Table 5. The fits to $S_t$ and $S_T$ include the different values of $k$ ($k=1$ and $k=2$). If the logarithmic argument in Eq. (\ref{stoscil}) were a precise description in the chemical systems, one would not expect the observed variation in parameters on going from $S_t$ to $S_T$. These variations are similar to the $a + b\ln N$ fits and show that the dependencies are more complex here than those in  Eq. (\ref{stoscil}).

\begin{table}[H]
	\begin{center}
		\begin{tabular}{c c c c c} 
			\hline
			$S_{t} (k=1)$ & $\alpha$ & $\lambda$ & $R^{2}_{adj}$ & $AICc$ \\ [0.5ex] 
			\hline
			$\mbox {Neutral atoms (He-Ar)}$ & -2.5329 & 0.9938 & 0.9998 & -53.4676 \\ 
			$\mbox {Homonuclear molecules (X-X)}$ & -2.6734 & 0.5205 & 0.9998 & -21.5225 \\
			$\mbox {Heteronuclear molecules (H-X)}$ & -2.7409 & 0.5361 & 0.9998 & -8.4899\\
			\hline 
			$S_{T}(k=2)$ & $\alpha$ & $\lambda$ & $R^{2}_{adj}$ & $AICc$ \\ [0.5ex] 
			\hline
			$\mbox {Neutral atoms (He-Ar)}$ & -4.4518 & 0.6364 & 0.9998 & -4.0145 \\ 
			$\mbox {Homonuclear molecules (X-X)}$ & -4.6326 & 0.2561 & 0.9998 & -1.8367 \\
			$\mbox {Heteronuclear molecules (H-X)}$ & -4.5327 & 0.3926 & 0.9998 & 0.8608 \\
			\hline
             $I_{T}$ & $\alpha$ & $\lambda$ & $R^{2}_{adj}$ & $AICc$ \\ [0.5ex] 
			\hline
			$\mbox {Neutral atoms (He-Ar)}$ & 1.0519 & 8.4006 & 0.9607 & -103.3061 \\ 
			$\mbox {Homonuclear molecules (X-X)}$ & 0.3897 & 3.0626 & 0.9961 & -156.9791 \\
			$\mbox {Heteronuclear molecules (H-X)}$ & 0.3119 & 2.3430 & 0.9691 & -59.8374 \\
			\hline		
		\end{tabular}
\caption{Values of $\alpha$ and $\lambda$ parameters obtained from fitting $S_t$, $S_T$ and $I_T$ data to Eqs. (\ref{stoscil}) and (\ref{mi}) functional forms. Fits $S_t$, $S_T$ and $I_T$ for the neutral atoms, homonuclear diatomic molecules and heteronuclear diatomic molecules are presented.}
	\end{center}
\end{table}
	
An analytical expression for the mutual information sum can be obtained from Eq. (\ref{stoscil}) by taking $2S^1_T-S^2_T$. This results in
\begin{equation}
I_T=\ln \bigg[\frac{[\omega(N-1)+\Lambda_{N}][\omega +(N-1)\Lambda_{N}]}{N^{2}\omega \Lambda_{N}} \bigg]-
 \frac{1}{2}\ln \bigg[\frac{[\omega(N-2)+2\Lambda_{N}][2\omega +(N-2)\Lambda_{N}]}{N^{2}\omega \Lambda_{N}} \bigg].
 \label{mi}
 \end{equation}
This form (with $\omega=N^{\alpha}$) can be fitted to the $I_T$ values. The parameters from this fit are given in Table 5.  One can appreciate how the $R^{2}_{adj}$ values from the fit to Eq. (\ref{mi}) are closer to unity than both of the previous models. It must be emphasized that any such comparison should be done with caution since this model contains non-linear parameters in contrast to the linear ones in the previous two models. However, this interpretation can also be corroborated from the $AICc$ metric. The values reported for $I_T$ present the largest negative values in the third model. Note that this third model is the only one where the expression for $I_T$ can be derived in a formal manner. The first two models for $I_T$ represent heuristic expressions obtained from empirical observations of the entropy sum expressions. It is also worth mentioning that the best performing fit here is that for the homonuclear diatomic molecules.

It is also significant that all three models provide fits to the entropic sums whose parameters do not vary widely over the three types of chemical species (atoms, homonuclear and heteronuclear diatomics) that were examined. These results add credence to the belief of a universal behaviour of the entropy sum, independent of the nature of the species that are considered.

\section{Conclusions} 
Pair density Shannon entropies are calculated at the Hartree-Fock level in both position and in momentum space in a series of homonuclear and heteronuclear diatomic molecules. The sum of these entropies form the basis of the entropic uncertainty relations. Mutual information, a statistical measure of correlation, is also discussed in these systems. The dependencies of these values on basis set is examined. 
The pair density Shannon entropy sum increases with basis set quality.
Mutual information in position space is seen to be larger than in momentum space for all studied systems. This extends the result, previously seen in neutral atoms, to molecular systems. The interpretation of this result is that the relative weight of the exchange component of the Hartree-Fock density is smaller in momentum space or that the distance from the Hartree-like reference is smaller in momentum space. 
The focus is also put on the $N$-dependent behaviour of the entropic and mutual information sums. Values of these sums are fitted to three different models in the three distinct chemical series. The three models are capable in varying degrees to represent the $N$-dependent behaviour of the entropic sums in the two diatomic series. These models also serve to model the behaviour of the pair density entropy sum.  The physical nature of the obtained parameters are discussed in the three models. These results add evidence to the conjecture that the entropy sum obeys a universal $N$-dependent relation independent of the nature of the particles that are present. Fits of the mutual information sum are not uniform among the models. The model whose origin is the ground state of interacting oscillators is shown to perform the best. The extension of such results to larger and more complex chemical species remains to be explored.

\section{Acknowledgments}
The authors thanks SECIHTI for the financial support S. J C. S. postdoctoral fellowship (CVU: 614012) and J.A.Z. postdoctoral fellowship (CVU: 667114) (Estancias posdoctorales por México).

\section{Data Availability Statement}
The data that support the findings of this study are available from the corresponding author upon reasonable request.

\newpage 
\appendix
\section{Computational details}\label{sec:compdutet}
\subsection{Electron density}
	The electron density, $\rho({\bf r})$, serves as the fundamental observable derived from quantum-chemical calculations. For molecular systems, it is typically expressed as a linear combination of occupied molecular orbitals $\chi_{m}({\bf r})$ \cite{schmidt1993}
    \begin{equation}
		\hat\rho({\bf r}) = \sum_{m=1}^{M} C_{m} \chi_{m}^{*}({\bf r}) \chi_{m}({\bf r}), 
	\end{equation}
	where $\hat\rho({\bf r})$ is the electron density (normalized to the number of electrons $N$, \textit{i.e.}, $\int\hat\rho({\bf r}) d{\bf r}=N$), $M$ is the number of occupied molecular orbitals, and $C_{m}$ are orbital
	occupation numbers. Throughout this section, densities with a hat (`$\hat{\phantom{x}}$') are normalized to the number of electrons ($N$) or the number of pairs $(N(N-1)/2)$, respectively. Each molecular orbital is constructed as a linear combination of atom-centered Gaussian-type primitive functions $\phi_{\dot{A}}(r)$	
	\begin{equation}
		\chi_{m}({\bf r}) = \sum_{\dot{A}=\dot{1}}^{\dot{P}} D_{m \dot{A}} \phi_{\dot{A}}({\bf r}- {{\bf R}}_{\dot{A}}),
	\end{equation}
	here, $\dot{P}$ denotes the total number of primitive functions, $D_{m \dot{A}}$ are the expansion coefficients, and ${\bf R}_{\dot{A}}$ are nuclear coordinates.
	The electron density can be calculated as
	\begin{equation}
		\hat\rho({\bf r})= \sum_{m=1}^{M}C_{m} \sum_{\dot{A}=\dot{1}}^{\dot P} \sum_{\dot{B}=\dot{1}}^{\dot P} D_{m {\dot A}} D_{m {\dot B}} \phi_{\dot{A}}({\bf r}) \phi_{\dot{B}}({\bf r}),
	\end{equation}
	this electron density can be rewritten compactly using a density matrix $c_{\dot{A} \dot{B}}$
	\begin{equation}\label{eq:rhodtk}
		\hat\rho({\bf r})= \phi_{\dot{A}}({\bf r}) c_{\dot{A} \dot{B}}  \phi_{\dot{B}}({\bf r}),
	\end{equation}
where the Einstein summation convention is implied. This representation is computationally efficient and is employed in widely used quantum-chemistry packages such as Gaussian 09 \cite{frisch2009}, as well as in density analysis tools like DensToolKit \cite{solano2024}.

Electron densities in momentum space are obtained throught the Fourier transform of the primitives $\phi({\bf r})_{\dot{A}}$, $\tilde\phi_{\dot{A}}({\bf p})$, which yield
\begin{equation}
    \hat\pi({\bf p})= \tilde\phi^*_{\dot{A}}({\bf p}) c_{\dot{A} \dot{B}}  \tilde\phi_{\dot{B}}({\bf p}).
\end{equation}

To recover Eqs.~(\ref{eq:shanonentropdef}), we use
\begin{equation}
    S_\rho=\frac{S_{\hat\rho}}{N}+\ln N, \qquad S_\pi=\frac{S_{\hat\pi}}{N}+\ln N.
\end{equation}

\subsection{Two-electron pair density}
The two-electron pair density, $\Gamma({\bf r}_{1}, {\bf r}_{2})$, describes the probability of simultaneously finding an electron at position ${\bf r_{1}}$ and another at ${\bf r_{2}}$. The spinless second order density is defined as
\begin{equation}
    \hat\Gamma_{HF}({{\bf r}}_{1},{{\bf r}}_{2}) = \frac{1}{2} \hat\rho({{\bf r}}_{1}) \hat\rho({{\bf r}}_{2}) -\frac{1}{2} \bigg[ \hat\Gamma_{x}^{\alpha,\alpha}({{\bf r}}_{1},{{\bf r}}_{2}) \hat\Gamma_{x}^{\alpha,\alpha}({{\bf r}}_{2},{{\bf r}}_{1}) + \hat\Gamma_{x}^{\beta,\beta}({{\bf r}}_{1},{{\bf r}}_{2}) \hat\Gamma_{x}^{\beta,\beta}({{\bf r}}_{2},{{\bf r}}_{1}) \bigg].
\end{equation}

For closed-shell systems with even number of pairwise electrons
\begin{equation}\label{eq:GHFspindepdef}
    \hat\Gamma_{x}^{\alpha,\alpha}({{\bf r}}_{1},{{\bf r}}_{2}) = \hat\Gamma_{x}^{\beta,\beta}({{\bf r}}_{1},{{\bf r}}_{2}) = \frac{1}{2} \hat\Gamma_{x}({{\bf r}}_{1},{{\bf r}}_{2}),
\end{equation}
therefore, Eq.~(\ref{eq:GHFspindepdef}) is reduced to \cite{solano2024,juan2000,parr1989}:
\begin{equation}
	\hat\Gamma_{HF}({{\bf r}}_{1}, {{\bf r}}_{2})= \frac{1}{2} \hat\rho({{\bf r}}_{1}) \hat\rho({{\bf r}}_{2}) -\frac{1}{4} \lvert \hat\Gamma_{x}({{\bf r}}_{1}, {{\bf r}}_{2}) \rvert^{2},
\end{equation}
where $\hat\Gamma_{x}({{\bf r}}_{1}, {{\bf r}}_{2})$ the one-particle reduced density, expressed in terms of primitive functions as
\begin{equation}
	\hat\Gamma_{x}({{\bf r}}_{1}, {{\bf r}}_{2})= \phi_{\dot{A}}({{\bf r}}_{1}) c_{\dot{A} \dot{B}} \phi_{\dot{B}}({{\bf r}}_{2}).
\end{equation}
For open-shell systems, the expression becomes more involved, requiring separate treatment of $\alpha$- and $\beta$-spin densities and density matrices \cite{solano2024}.

Similar expressions are obtained in momentum space, by replacing $\phi_{\dot{A}}({\bf r})$ with $\tilde\phi_{\dot{A}}({\bf p})$. For instance, for closed-shell systems
\begin{equation}
    \hat\Pi_{HF}({{\bf p}}_{1}, {{\bf p}}_{2})= \frac{1}{2} \hat\pi({{\bf p}}_{1}) \hat\pi({{\bf p}}_{2}) -\frac{1}{4} \lvert \hat\Pi_{x}({{\bf p}}_{1}, {{\bf p}}_{2}) \rvert^{2},
\end{equation}
where
\begin{equation}
	\hat\Pi_{x}({{\bf p}}_{1}, {{\bf p}}_{2})= \tilde\phi^*_{\dot{A}}({{\bf p}}_{1}) c_{\dot{A} \dot{B}} \tilde\phi_{\dot{B}}({{\bf p}}_{2}).
\end{equation}

The integration of $\hat\Gamma_{HF}$ over six-dimensional space $\Omega \times \Omega$ yields the total number of electron pairs
\begin{equation}\label{eq:integhatGammadtk}
	\int_{\Omega \times \Omega} \hat\Gamma_{HF}({{\bf r}}_{1}, {{\bf r}}_{2}) \mathrm{d} {{\bf r}}_{1} \mathrm{d} {{\bf r}}_{2} = \frac{N(N-1)}{2},
\end{equation}
where $N$ is the total number of electrons.

To recover Eqs.~(\ref{eq:pairentropiesdef}), we use
\begin{equation}\label{eq:unnorm2normS2s}
   S_\Gamma=\frac{2}{N(N-1)}S_{\hat\Gamma}+\ln\left(\frac{N(N-1)}{2}\right),
   \qquad
   S_\Pi=\frac{2}{N(N-1)}S_{\hat\Pi}+\ln\left(\frac{N(N-1)}{2}\right).
\end{equation}

\section{Numerical integration schemes for diatomic systems}\label{sec:numintschs}
	
A cubature rule provides an approximation to a three-dimensional integral through a discrete sum
\begin{equation}
	I = \int_{\Omega} f({{\bf r}})  \mathrm{d} {\bf r} \approx \sum_{i} \omega_{i} f\big({\bf r_{i}} \big),
\end{equation}
where $\{w_{i} \}$ and $\{\bf{ r_{i}} \}$ are weights and abscissae, respectively. The accuracy of the scheme depends on the choice of coordinate transformation and the underlying one-dimensional quadrature rules for each independent variable.

\subsection{Specialized scheme for diatomic molecules}
For diatomic systems, we employ a tailored integration scheme that exploits the azimuthal symmetry of the electron density. The spatial domain $\Omega=\mathbb{R}^{3},$ is decomposed into four subregions, two core spherical regions centered on each nucleus, and two valence regions that occupy the remaining space between the nuclei.
	
\begin{figure}[h!]
	\begin{center}
	\includegraphics[width=0.5 \linewidth]
	{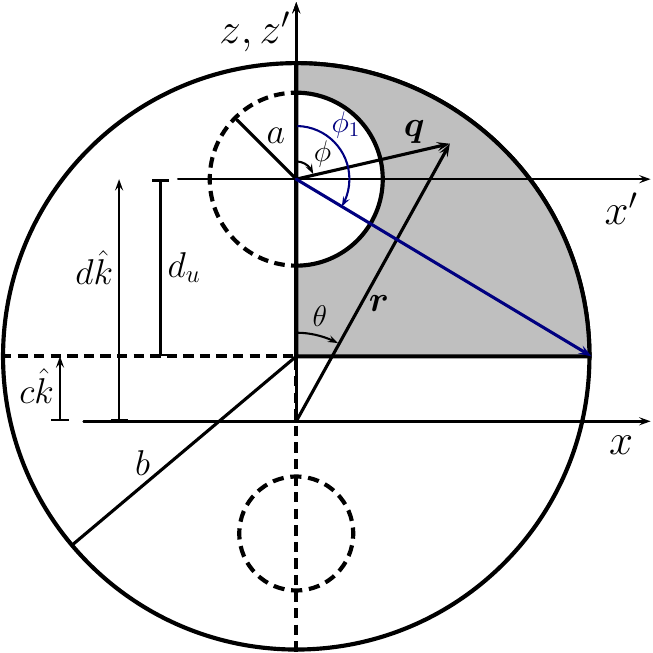}
	\caption{The azimuthally symmetric upper half of the domain (denoted $\Sigma_{2v}^{u}$) is mapped using a polar coordinate system $(q,\phi)$.} 
	\label{figure2}
	\end{center}
\end{figure}

The valence regions are further divided into upper and lower halves. Using a coordinate transformation, the integration over a valence subregion reduces to
\begin{multline}
	\int_{\Sigma2v} f(r,\theta) \mathrm{d}V = 2\pi \int_{0}^{\phi_{1}} \sin \phi \mathrm{d}\phi \int_{a}^{B(\phi)} q^{2}f(q,\phi) \mathrm{d}q + \\
	2\pi \int_{\phi_{1}}^{\pi} \sin \phi \mathrm{d}\phi \int_{a}^{C(\phi)} q^{2}f(q,\phi) \mathrm{d}q,
\end{multline}
where $B(\phi)$ and $C(\phi)$ are limits defined by the geometry of the diatomic system, see Figure \ref{figure2}
\begin{align*}
	B(\phi) &\equiv \sqrt{b^{2}-d^{2}_{u} \sin^{2} \phi} -d_{u}\cos \phi, \\
	C(\phi) &\equiv -\frac{d_{u}}{\cos \phi}.
\end{align*}
	
\begin{figure}[h!]
	\begin{center}
	\includegraphics[width=0.5 \linewidth]
	{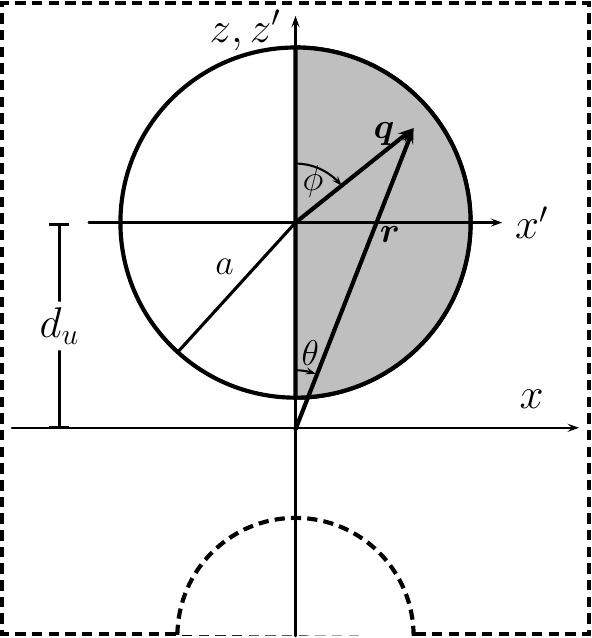}
	\caption{The upper core part of the integration domain (denoted as $\Sigma^u_{2c}$ in the text), and its associated local coordinates $\phi$ and $q$. Here $d_u$ is the distance from the global origin to the upper nucleus.} 
	\label{figure3}
	\end{center}
\end{figure}

Core region integration (shown in Figure \ref{figure3}) treats the spherical domains centered on each nucleus (denoted $\Sigma_{2c}^{u}$ and $\Sigma_{2c}^{l}$) using standard spherical coordinates. The total integration domain for a diatomic molecule is thus
\begin{equation}
    \Sigma_{2}= \Sigma_{2v}^{u} \cup \Sigma_{2c}^{u} \cup \Sigma_{2v}^{l} \cup \Sigma_{2c}^{l},
\end{equation}
If the system lacks azimuthal symmetry (e.g., open-shell molecules), an additional quadrature is applied for the azimuthal angle $\phi$. All radial and angular integrations are performed using standard Gauss–Legendre quadrature. These numerical integration schemes have been used \cite{solano_2025}.
	
The integration domain is bounded by selecting a radius $b$ such that $\rho(b \hat{z}) \leq 10^{-14}$ a.u., ensuring the integrand is negligible at the boundary. The core–valence boundary is set at 0.75 times the van der Waals radius of the respective atom. 

To preserve numerical consistency, we integrated Eqs.~(\ref{eq:rhodtk}) and
(\ref{eq:integhatGammadtk}) to
obtain $N$, and $N(N-1)/2$, respectively. Subsequently, we used these numerically integrated values to compute $S_{\Gamma}$ and $I_r$.
	
\subsection{Extension to two-electron densities}
The two-electron pair density $\Gamma({\bf r}_{1},{\bf r}_{2})$ is a function of six variables. Even for diatomics with azimuthal symmetry in 3D, this symmetry is generally lost in the 6D space $\Omega \times \Omega$, so no straightforward reduction in the number of integration variables is available.
We extend the diatomic 3D scheme by constructing a product cubature rule 
\begin{equation}
	\int_{\Omega \times \Omega} P({\bf r}_{1},{\bf r}_{2}) \mathrm{d} {\bf r}_{1} \mathrm{d} {\bf r}_{2} \approx \sum_{i,j} w_{i} w_{j} P({\bf r}_{1},{\bf r}_{2}),
\end{equation}
where $\{ w_{i}, {\bf {r}}_{i} \}$ and $\{ w_{j}, {\bf {r}}_{j} \}$ are the weights and abscissas from two independent 3D diatomic cubature rules. This requires explicit integration over all angular variables, including azimuthal angles $\phi_{1}$ and $\phi_{2}$, even for symmetric molecules.

Integrals in position space were computed using cubatures composed of 32, 32, and 9 abscissas for the radial, azymuthal, and polar coordinates, which rendered a total of 27\,648 3D integration points and 764\,411\,904 6D points. The above combination produces one-electron density integrals with relative errors between 3.247$\times10^{-8}$ and 8.985$\times10^{-5}$, and two-electron density integrals with relative errors between 6.494$\times10^{-8}$ to 1.797$\times10^{-4}$.

\subsection{Numerical integration schemes for momentum space}	
In momentum space, 3D functions are integrated within a spherical domain of radius $Q$. We choose $Q$ in such a manner that
\[
    \text{max}\left(\hat\pi(Q\hat\imath),\hat\pi(Q\hat\jmath),
    \hat\pi(Q\hat k)\right)\leq10^{-12}\text{a.u.}
\]
Cubature rules were constructed using Gauss-Legendre quadratures for the radial ($p$) coordinate and spherical-t designs \cite{bib:an-xiao2019} for the solid angles.

Integrals in momentum space were computed using 96 Gauss-Legendre integration points and a spherical t-design of order 13, which yield 9\,024 3D points and 81\,432\,576 6D points. Such combination produces integrals of one-electron momentum densities with relative errors between 3.247$\times10^{-8}$ and
 8.985$\times10^{-5}$, and two-electron momentum density with relative errors between 1.30$\times10^{-8}$ and  2.137$\times10^{-3}$.

\section{Neutral atoms and heteronuclear diatomic molecules}
\label{sec:appendixc}

	\begin{figure}[H]
		\includegraphics[width=\linewidth]
		{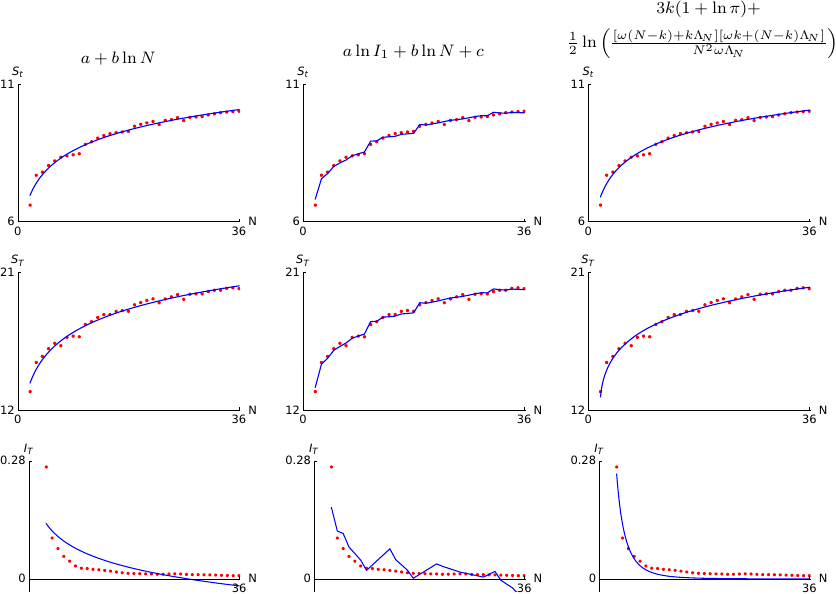}
		\caption{Shannon entropy sums $S_t$ (first row), $S_T$ (second row) and $I_T$ (third row) as a function of $N$ for the neutral atoms. Columns correspond to the first [Eq. (\ref{gadre-st})], second [Eq. (\ref{ent-n-ion})] and third [Eq. (\ref{stoscil})] models respectively. The red dots are the actual data points obtained with the 6-31 G basis set while the blue continuous curves are the respective fits. }
		\label{fig4}
	\end{figure}

	\begin{figure}[H]
		\includegraphics[width=\linewidth]
		{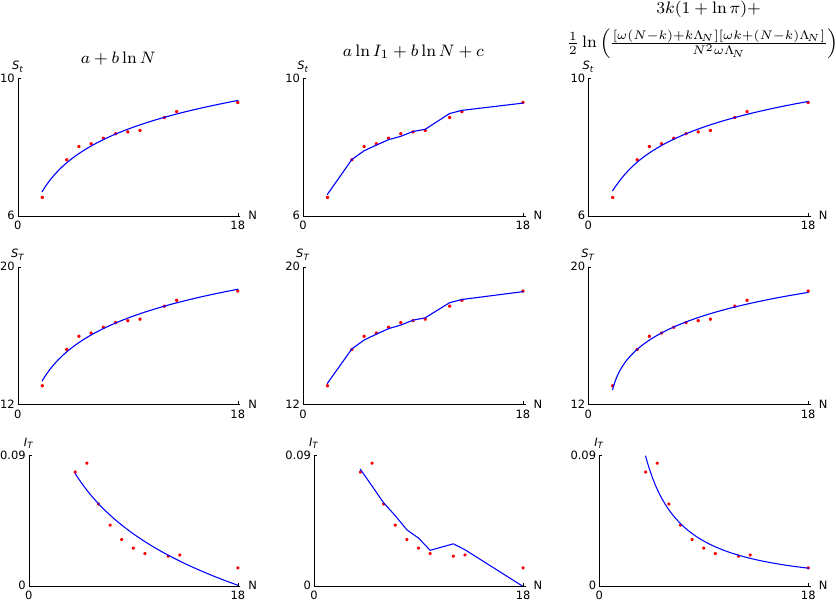}
		\caption{Shannon entropy sums $S_t$ (first row), $S_T$ (second row) and $I_T$ (third row) as a function of $N$ for the heteronuclear diatomic molecules. Columns correspond to the first [Eq. (\ref{gadre-st})], second [Eq. (\ref{ent-n-ion})] and third [Eq. (\ref{stoscil})] models respectively. The red dots are the actual data points obtained with the 6-31 G basis set while the blue continuous curves are the respective fits. }
		\label{fig5}
	\end{figure}	
\newpage
\bibliographystyle{unsrt}
\bibliography{bibliography}

\end{document}

%% file: tab-hxseries-hf-diffbasis.tex
\begin{footnotesize}
	\begin{longtable}{clrrrrrr}
	\hline\hline
   & Basis set & $S_{\Gamma}$ & $S_{\Pi}$ & $S_{T}$
& $10^2\cdot I_{r}$ & $10^2\cdot I_{p}$ & $10^2\cdot I_{T}$ \\\hline
H-H & STO-3G & 7.6442 & 5.4200 & 13.0642 & 0.0000 & 0.0000 & 0.0000\\
 & 3-21G & 7.8593 & 5.2025 & 13.0618 & 0.0000 & 0.0000 & 0.0000\\
 & 6-31G & 7.8985 & 5.1832 & 13.0816 & 0.0000 & 0.0000 & 0.0000\\
 & 6-31G(2df,p) & 7.8748 & 5.2095 & 13.0844 & 0.0000 & 0.0000 & 0.0000\\
 \\
H-Li & STO-3G & 7.3019 & 7.8991 & 15.2009 & 5.0309 & 2.6384 & 7.6693\\
 & 3-21G & 7.4382 & 7.7639 & 15.2021 & 5.1690 & 2.6942 & 7.8632\\
 & 6-31G & 7.4347 & 7.7599 & 15.1945 & 5.1272 & 2.7147 & 7.8419\\
 & 6-31G(2df,p) & 7.4278 & 7.7554 & 15.1832 & 5.1241 & 2.6810 & 7.8051\\
 \\
H-Be & STO-3G & 7.2440 & 8.6739 & 15.9179 & 5.4957 & 2.6333 & 8.1289\\
 & 3-21G & 7.6331 & 8.3477 & 15.9808 & 5.7945 & 2.6827 & 8.4771\\
 & 6-31G & 7.5632 & 8.3991 & 15.9623 & 5.7401 & 2.7129 & 8.4530\\
 & 6-31G(2df,p) & 7.5442 & 8.4043 & 15.9486 & 5.8522 & 2.6935 & 8.5456\\
 \\
H-B & STO-3G & 6.7295 & 9.3659 & 16.0953 & 3.1752 & 2.4228 & 5.5981\\
 & 3-21G & 7.2593 & 8.9063 & 16.1656 & 3.2210 & 2.4235 & 5.6444\\
 & 6-31G & 7.1857 & 8.9623 & 16.1480 & 3.1790 & 2.4622 & 5.6412\\
 & 6-31G(2df,p) & 7.1780 & 8.9721 & 16.1502 & 3.2615 & 2.4292 & 5.6907\\
 \\
H-C & STO-3G & 6.3806 & 10.0558 & 16.4363 & 2.4179 & 1.8066 & 4.2245\\
 & 3-21G & 6.7294 & 9.7520 & 16.4814 & 2.3807 & 1.8104 & 4.1911\\
 & 6-31G & 6.7096 & 9.7773 & 16.4869 & 2.3622 & 1.8246 & 4.1868\\
 & 6-31G(2df,p) & 6.7006 & 9.7866 & 16.4872 & 2.4299 & 1.7984 & 4.2283\\
 \\
H-N & STO-3G & 5.9622 & 10.7293 & 16.6915 & 1.8969 & 1.4073 & 3.3042\\
 & 3-21G & 6.1440 & 10.5798 & 16.7238 & 1.8394 & 1.3722 & 3.2116\\
 & 6-31G & 6.1878 & 10.5706 & 16.7584 & 1.8193 & 1.3778 & 3.1971\\
 & 6-31G(2df,p) & 6.1750 & 10.5799 & 16.7550 & 1.8437 & 1.3710 & 3.2147\\
 \\
H-O & STO-3G & 5.3786 & 11.4101 & 16.7886 & 1.5321 & 1.1553 & 2.6874\\
 & 3-21G & 5.5678 & 11.2580 & 16.8257 & 1.4718 & 1.1369 & 2.6087\\
 & 6-31G & 5.6314 & 11.2392 & 16.8707 & 1.4541 & 1.1494 & 2.6034\\
 & 6-31G(2df,p) & 5.6182 & 11.2471 & 16.8653 & 1.4713 & 1.1465 & 2.6178\\
 \\
H-F & STO-3G & 4.8356 & 12.0430 & 16.8786 & 1.3042 & 0.9949 & 2.2991\\
 & 3-21G & 5.0119 & 11.9048 & 16.9167 & 1.2525 & 0.9807 & 2.2332\\
 & 6-31G & 5.0725 & 11.8865 & 16.9590 & 1.2371 & 0.9915 & 2.2286\\
 & 6-31G(2df,p) & 5.0637 & 11.8901 & 16.9538 & 1.2334 & 0.9958 & 2.2292\\
 \\
H-Na & STO-3G & 4.8077 & 12.7018 & 17.5095 & 1.1981 & 0.7650 & 1.9631\\
 & 3-21G & 5.1460 & 12.5568 & 17.7027 & 1.2347 & 0.8369 & 2.0716\\
 & 6-31G & 5.1585 & 12.5566 & 17.7150 & 1.2131 & 0.8353 & 2.0485\\
 & 6-31G(2df,p) & 5.1487 & 12.5611 & 17.7099 & 1.2141 & 0.8340 & 2.0481\\
 \\
H-Mg & STO-3G & 4.8814 & 12.9679 & 17.8493 & 1.1552 & 0.7735 & 1.9288\\
 & 3-21G & 5.3456 & 12.7047 & 18.0503 & 1.3176 & 0.8273 & 2.1448\\
 & 6-31G & 5.3500 & 12.7065 & 18.0565 & 1.3031 & 0.8291 & 2.1322\\
 & 6-31G(2df,p) & 5.3430 & 12.7125 & 18.0555 & 1.3213 & 0.8265 & 2.1478\\
 \\
H-Cl & STO-3G & 4.7525 & 13.7799 & 18.5324 & 0.7056 & 0.5678 & 1.2734\\
 & 3-21G & 4.8814 & 13.7137 & 18.5951 & 0.6792 & 0.5697 & 1.2489\\
 & 6-31G & 4.8731 & 13.7220 & 18.5951 & 0.6738 & 0.5697 & 1.2435\\
 & 6-31G(2df,p) & 4.8645 & 13.7267 & 18.5912 & 0.6762 & 0.5696 & 1.2458\\
 \hline\hline
 \caption{Entropies and mutual informations for the heteronuclear series $H-X$ obtained with different basis functions.}
	\end{longtable}
\end{footnotesize}

%% file: tab-xxneut-hf-diffbasis.tex
\begin{footnotesize}
\begin{longtable}{clrrrrrr}
	\hline
   & Basis set & $S_{\Gamma}$ & $S_{\Pi}$ & $S_{T}$
& $10^2\cdot I_{r}$ & $10^2\cdot I_{p}$ & $10^2\cdot I_{T}$ \\\hline
Li-Li & STO-3G & 7.9222 & 8.2597 & 16.1819 & 3.8223 & 2.5396 & 6.3619\\
 & 3-21G & 8.3310 & 7.9257 & 16.2567 & 3.9604 & 2.5794 & 6.5398\\
 & 6-31G & 8.2545 & 7.9977 & 16.2523 & 3.9065 & 2.6364 & 6.5428\\
 & 6-31G(2df,p) & 8.2535 & 8.0008 & 16.2543 & 3.9207 & 2.6418 & 6.5625\\
 \\
Be-Be & STO-3G & 7.8760 & 8.9579 & 16.8339 & 2.5768 & 1.9028 & 4.4796\\
 & 3-21G & 8.3559 & 8.5228 & 16.8787 & 2.6319 & 1.9483 & 4.5802\\
 & 6-31G & 8.2640 & 8.6045 & 16.8685 & 2.5956 & 1.9767 & 4.5723\\
 & 6-31G(2df,p) & 8.2740 & 8.5998 & 16.8737 & 2.5980 & 1.9835 & 4.5814\\
 \\
B-B & STO-3G & 7.2132 & 9.9764 & 17.1896 & 1.7314 & 1.2690 & 3.0004\\
 & 3-21G & 7.7863 & 9.4797 & 17.2660 & 1.7599 & 1.3181 & 3.0780\\
 & 6-31G & 7.7044 & 9.5400 & 17.2445 & 1.7362 & 1.3395 & 3.0757\\
 & 6-31G(2df,p) & 7.7080 & 9.5348 & 17.2428 & 1.7490 & 1.3367 & 3.0857\\
 \\
C-C & STO-3G & 6.8326 & 10.5670 & 17.3997 & 1.2648 & 0.9122 & 2.1769\\
 & 3-21G & 7.1372 & 10.3362 & 17.4734 & 1.2475 & 0.9096 & 2.1570\\
 & 6-31G & 7.1033 & 10.3662 & 17.4695 & 1.2349 & 0.9231 & 2.1580\\
 & 6-31G(2df,p) & 7.0922 & 10.3616 & 17.4537 & 1.2401 & 0.9256 & 2.1658\\
 \\
N-N & STO-3G & 6.4105 & 11.2155 & 17.6260 & 1.0452 & 0.7617 & 1.8069\\
 & 3-21G & 6.5270 & 11.1590 & 17.6860 & 1.0154 & 0.7293 & 1.7448\\
 & 6-31G & 6.5514 & 11.1644 & 17.7158 & 1.0074 & 0.7324 & 1.7398\\
 & 6-31G(2df,p) & 6.5331 & 11.1754 & 17.7085 & 1.0273 & 0.7250 & 1.7523\\
 \\
O-O & STO-3G & 5.9670 & 11.8592 & 17.8262 & 0.9092 & 0.6080 & 1.5172\\
 & 3-21G & 6.1736 & 11.7294 & 17.9030 & 0.8859 & 0.5987 & 1.4846\\
 & 6-31G & 6.2069 & 11.7413 & 17.9482 & 0.8769 & 0.6023 & 1.4792\\
 & 6-31G(2df,p) & 6.1984 & 11.7433 & 17.9417 & 0.8846 & 0.6036 & 1.4882\\
 \\
F-F & STO-3G & 5.5463 & 12.4406 & 17.9869 & 0.8386 & 0.4886 & 1.3272\\
 & 3-21G & 5.7827 & 12.2991 & 18.0818 & 0.8133 & 0.4864 & 1.2997\\
 & 6-31G & 5.8160 & 12.3073 & 18.1233 & 0.8037 & 0.4885 & 1.2922\\
 & 6-31G(2df,p) & 5.8070 & 12.3067 & 18.1137 & 0.8032 & 0.4911 & 1.2943\\
 \\
Ne-Ne & STO-3G & 5.0614 & 13.0786 & 18.1400 & 0.7561 & 0.3262 & 1.0823\\
 & 3-21G & 5.3817 & 12.8933 & 18.2750 & 0.7362 & 0.3314 & 1.0676\\
 & 6-31G & 5.4424 & 12.8823 & 18.3247 & 0.7299 & 0.3439 & 1.0738\\
 & 6-31G(2df,p) & 5.4365 & 12.8829 & 18.3194 & 0.7289 & 0.3444 & 1.0732\\
 \\
Na-Na & STO-3G & 5.5747 & 13.1390 & 18.7137 & 0.6197 & 0.3323 & 0.9520\\
 & 3-21G & 5.9439 & 12.9355 & 18.8794 & 0.6938 & 0.3743 & 1.0682\\
 & 6-31G & 5.9515 & 12.9396 & 18.8910 & 0.6804 & 0.3857 & 1.0661\\
 & 6-31G(2df,p) & 5.9508 & 12.9404 & 18.8912 & 0.6815 & 0.3855 & 1.0670\\
 \\
Mg-Mg & STO-3G & 5.6502 & 13.3656 & 19.0158 & 0.6009 & 0.3540 & 0.9549\\
 & 3-21G & 6.1326 & 13.0486 & 19.1811 & 0.6377 & 0.3824 & 1.0200\\
 & 6-31G & 6.1321 & 13.0392 & 19.1713 & 0.6300 & 0.3818 & 1.0118\\
 & 6-31G(2df,p) & 6.1330 & 13.0380 & 19.1710 & 0.6294 & 0.3820 & 1.0114\\
 \\
Al-Al & STO-3G & 5.6536 & 13.5742 & 19.2278 & 0.5627 & 0.3279 & 0.8906\\
 & 3-21G & 6.1466 & 13.2433 & 19.3899 & 0.5468 & 0.3274 & 0.8742\\
 & 6-31G & 6.1492 & 13.2473 & 19.3965 & 0.5387 & 0.3337 & 0.8723\\
 & 6-31G(2df,p) & 6.1490 & 13.2494 & 19.3984 & 0.5414 & 0.3325 & 0.8739\\
 \\
Si-Si & STO-3G & 5.7173 & 13.7253 & 19.4426 & 0.5013 & 0.3005 & 0.8017\\
 & 3-21G & 6.0663 & 13.4763 & 19.5426 & 0.4974 & 0.3119 & 0.8093\\
 & 6-31G & 6.0632 & 13.4776 & 19.5407 & 0.4924 & 0.3135 & 0.8059\\
 & 6-31G(2df,p) & 6.0633 & 13.4819 & 19.5452 & 0.5001 & 0.3100 & 0.8101\\
 \\
P-P & STO-3G & 5.6498 & 13.9126 & 19.5624 & 0.4405 & 0.2663 & 0.7067\\
 & 3-21G & 5.9455 & 13.6996 & 19.6450 & 0.4340 & 0.2719 & 0.7059\\
 & 6-31G & 5.9400 & 13.7044 & 19.6444 & 0.4302 & 0.2751 & 0.7054\\
 & 6-31G(2df,p) & 5.9290 & 13.7075 & 19.6364 & 0.4366 & 0.2719 & 0.7085\\
 \\
S-S & STO-3G & 5.5775 & 14.0970 & 19.6746 & 0.4185 & 0.2477 & 0.6661\\
 & 3-21G & 5.8433 & 13.9118 & 19.7551 & 0.4085 & 0.2524 & 0.6608\\
 & 6-31G & 5.8343 & 13.9189 & 19.7532 & 0.4053 & 0.2547 & 0.6600\\
 & 6-31G(2df,p) & 5.8358 & 13.9148 & 19.7506 & 0.4100 & 0.2549 & 0.6649\\
 \\
Cl-Cl & STO-3G & 5.6272 & 14.1511 & 19.7784 & 0.4058 & 0.2348 & 0.6406\\
 & 3-21G & 5.7258 & 14.1111 & 19.8369 & 0.3912 & 0.2348 & 0.6260\\
 & 6-31G & 5.7132 & 14.1206 & 19.8339 & 0.3885 & 0.2350 & 0.6234\\
 & 6-31G(2df,p) & 5.7149 & 14.1133 & 19.8282 & 0.3894 & 0.2368 & 0.6262\\
 \\
Ar-Ar & STO-3G & 5.4817 & 14.3978 & 19.8795 & 0.3843 & 0.1880 & 0.5723\\
 & 3-21G & 5.6550 & 14.3028 & 19.9578 & 0.3728 & 0.1987 & 0.5715\\
 & 6-31G & 5.6433 & 14.3108 & 19.9541 & 0.3703 & 0.1918 & 0.5621\\
 & 6-31G(2df,p) & 5.6432 & 14.3101 & 19.9533 & 0.3702 & 0.1967 & 0.5669\\
 \\
 \hline\hline
 \caption{Entropies and mutual informations for the homonuclear series $X-X$ obtained with different basis functions.}
\end{longtable}
\end{footnotesize}